\documentclass[a4paper,prl,twocolumn]{revtex4}
\usepackage{graphicx}
\usepackage{subfigure}
\usepackage{amssymb}
\usepackage{xcolor}
\usepackage{soul}           

\begin{document}
\title{Electron inelastic mean free paths in condensed matter down to a few electronvolts}

\author{Pablo de Vera}

\email[Corresponding author: ]{pablo.vera@um.es}


\author{Rafael Garcia-Molina}

\affiliation{Departamento de F\'{i}sica -- Centro de Investigaci\'{o}n en \'{O}ptica y Nanof\'{i}sica,  Regional Campus of International Excellence ``Campus Mare Nostrum'', Universidad de Murcia, E-30100 Murcia, Spain}


\begin{abstract}

A method is reported for a simple, yet reliable, calculation of electron inelastic mean free paths in condensed phase insulating and conducting materials, from the very low energies of hot electrons up to the high energies characteristic of electron beams.  Through a detailed consideration of the energy transferred by the projectile in individual and collective electronic excitations, as well as ionizations, together with the inclusion of higher order corrections to the results provided by the dielectric formalism,  inelastic mean free paths are calculated for water, aluminum, gold and copper in excellent agreement with the available experimental data, even at the elusive very low energy region. 
These results are important due to the crucial role played by low energy electrons in radiobiology (owing to their relevant effects in biodamage), and also in order to assess the not yet elucidated disagreement between older and recent measurements of low energy electron mean free paths in metals (which are relevant for low energy electron transport and effects in nanostructured devices).


\end{abstract}

\maketitle

Electrons interacting (in a broad range of energies) with condensed matter appear either as direct projectiles, as the result of target ionization by external radiation, or as charge carriers in electronic devices. 
The precise knowledge of the transport of these electrons through matter, which depends on the electron energy and the medium characteristics, provides very useful information for controlling material properties (through modification or analysis) \cite{Egerton2011,Dapor2017}, improving the yield of nanoelectronic devices that relies on charge mobility \cite{Cheng2005,Lundstrom2006}, or to gain knowledge on energy conversion, catalysis at surfaces and nanomaterials \cite{Brown2016,Hartland2017} or oncological studies  at the molecular level \cite{Solov'yov2017,Zheng2018}, all of them depending on the transport and intereactions of electrons.
To properly understand and model all these phenomena, the knowledge of the average distance between inelastic collisions (i.e., the inelastic mean free path, IMFP) is of paramount relevance.


The dielectric framework to treat the interaction of charged particles with matter, which dates back to Fermi \cite{Fermi1940}, relies on the first Born approximation (FBA).
Currently, it represents a reliable, yet simple, method to calculate electronic mean free paths \cite{Nikjoo2012} and, even, ionization cross sections in condensed matter \cite{deVera2013,deVera2015}. General consensus between models and experimental data is found for electron energies $\gtrsim 200$ eV \cite{Powell1999,Da2014}, with discrepancies appearing at the lower energies ($\lesssim 50$ eV). 
This raises some debate \cite{Emfietzoglou2005,NguyenTruong2016,NguyenTruong2017}
due to the influence of the maximum energy transferred by the electron when interacting with the target \cite{Denton2008,Bourke2012,NguyenTruong2013}. Also, due to possible corrections to the FBA for low energy (sometimes referred as ``hot'') electrons \cite{Emfietzoglou2013},
so relevant in nanoelectronics \cite{Lundstrom2006}, catalysis \cite{Hartland2017} or biomolecular damage \cite{Solov'yov2017,Zheng2018}.


In this letter we discuss the different role played by (both individual and collective) excitations, as well as ionizations, induced by electrons in insulators and conductors, as they have rather different types of electronic excitations. This is illustrated for representative materials (water, aluminum, gold and copper) relevant for radiological and nanoelectronic applications.  Water is a molecular material where collective electronic excitations are unlikely \cite{Hayashi2015}, thus inelastic interactions result in ionized (i.e., free) or excited (i.e., bound) electrons. In metals, both collective (e.g., plasmons) and individual electronic excitations are possible, the latter directly leading to free electrons in the conduction band.
While the excitation spectrum of aluminum is dominated by a strong plasmon, gold and copper present complex excitation spectra, where both individual and collective excitations can coexist.

In this work, we will show that a proper consideration of the excitation spectrum of the target, together with the inclusion of higher order corrections to the FBA, allow the calculation of IMFP in excellent agreement with the available experimental data for a wide energy range, down to a few electronvolts. In particular, for these materials (especially for metals) there is an extensive set of experimental data over a wide energy range which will serve as a benchmark for our model. For the case of copper, there is also a discrepancy between
the low energy IMFP
derived from modern experimental techniques \cite{Bourke2010} 
and older measurements \cite{Knapp1979,Ogawa1997,Bauer2015}. Our analysis will provide clues in order to elucidate these discrepancies.

Within the dielectric formalism (i.e., FBA), the inelastic mean free path $\lambda_{\rm e}$ of an electron of kinetic energy $T$, mass $m$ and charge $e$ is  \cite{Nikjoo2012,GarciaMolina2017}:
\begin{eqnarray}
\lambda_{\rm e}^{-1}(T) & = & \frac{e^2m}{\pi \hbar^2 T} \sum_{i=1}^n \int_{E_{-,i}}^{E_{+,i}} {\rm d}E \int_{k_-}^{k_+} \frac{{\rm d}k}{k} f_{\rm ex}(k,T) \nonumber \\ & & \times F(E-E_{{\rm th},i}){\rm Im}\left[ \frac{-1}{\epsilon(k,E)} \right]_i ,
\label{eq:IMFP}
\end{eqnarray}
where $E=\hbar\omega$ and $\hbar k$ are, respectively, the energy and momentum transferred in an inelastic collision. The target electronic excitation spectrum is represented by its Energy-Loss Function (ELF), ${\rm Im}\left[ \frac{-1}{\epsilon(k,E)} \right] = \sum_i F(E-E_{{\rm th},i}){\rm Im}\left[ \frac{-1}{\epsilon(k,E)} \right]_i$,
with $F(E-E_{{\rm th},i})$ and $E_{{\rm th},i}$ being a smooth step function and threshold energies, respectively. Although the contributions $i$ are commonly used to reproduce the measured ELF of materials, they can be given a physical meaning, i.e., the excitation of different electronic levels, as it will be explained later on. The indistinguishability between the incident and target electrons (when applicable) is introduced through the Born-Ochkur exchange factor $f_{\rm ex}(k,T)$ \cite{FernandezVarea1993}. 

In practice, Eq.(\ref{eq:IMFP}) is evaluated by means of extended optical data models \cite{Garcia-Molina2012}, where the ELF is usually fitted to experimental data in the optical limit ($k=0$), and appropriately extended to finite momentum transfers (i.e., over the whole Bethe surface). In this work, we use the MELF-GOS method to  cover the whole $(k,E)$-space \cite{Garcia-Molina2012}. 

The IMFP calculation depends, as seen in Eq.(\ref{eq:IMFP}), on the integration limits on momentum and energy transfers.
The former, obtained by energy and momentum conservation, are $\hbar k_{\pm} = \sqrt{2m}\left( \sqrt{T} \pm \sqrt{T-E} \right)$. The latter, generally considered independent of the electronic excitation ($E_{\pm,i} = E_{\pm}$) are determined not only by energy conservation, but also by the Pauli exclusion principle and electron indistinguishability. On the one hand, the primary electron cannot fall into an occupied level of the target. Therefore, for metals, 
$E_{+} = T-E_{\rm F}$, where $E_{\rm F}$ is the Fermi energy. On the other hand, when a primary electron creates a free secondary electron (previously bound with energy $B_i$), 
the maximum energy transfer 
occurs when 
both have the same final energy, so $E_{+} = (T+B_i)/2$. 
On this basis, Bourke and Chantler \cite{Bourke2012} discussed whether the maximum energy transfer (in the case of metals) should be $\sim T/2$ or simply $T-E_{\rm F}$. They concluded that the former is a too constraining limit, arguing that collective excitations (e.g., plasmons, distinguishable from the primary electron), represent the main excitation channel in metals. 
Considering the different nature of each inelastic interaction will lead to maximum energy transfers that depend on the specific electronic excitation ($E_{\pm,i} \neq E_{\pm}$), as explained in what follows.

For metals, both individual and collective electronic excitations (e.g., plasmons) are possible. Every individual transition will promote an electron to the partially filled conduction band (through which the primary electron is moving), 
so $E_{+,i}^{\rm indiv} = {\rm min}\left[(T+B_i)/2,T-E_{\rm F}\right]$. In turn, a collective excitation is distinguishable from the primary electron, thus 
$E_{+,i}^{\rm collect} = T-E_{\rm F}$ and $f_{\rm ex} = 0$. 

For insulators, two possible types of excitations will be considered: an electron transition to a localized discrete energy level (an ``excitation''), or to the conduction band (an ``ionization''). 
As shown in Refs. \cite{deVera2013,deVera2015} for water 
 and other molecular materials, the introduction of a mean binding energy $\overline{B}$ for the outer shell electrons allows the distinction between both types of excitations. When $E < \overline{B}$, an electron is excited to a bound state, while the primary electron moves freely in the conduction band; the latter can then lose all its energy, 
and $E_{+,i}^{\rm excit} = {\rm min}\left[T,\overline{B}\right]$. If $E > \overline{B}$, an electron is ionized; then both primary and secondary electrons move in the conduction band,
so $E_{+,i}^{\rm ioniz} = (T+\overline{B})/2$. All these criteria will be assessed later on, when comparing the calculated IMFP with experimental data.
 

\begin{figure}[t]
	\centering
	\includegraphics[width=1.0\columnwidth]{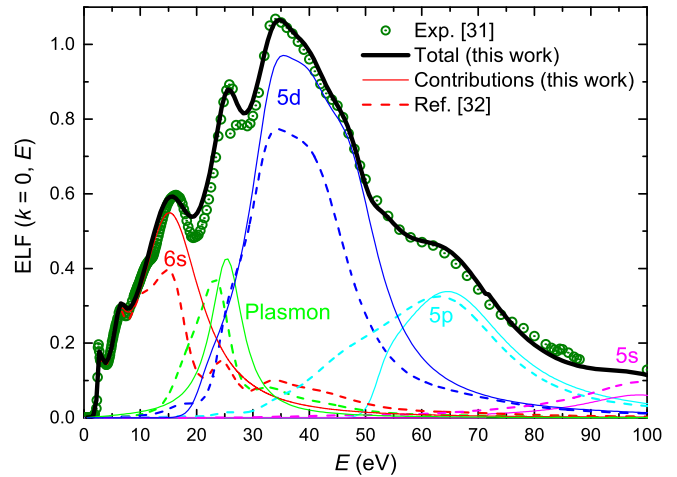}
	\caption{(Color online) Energy-loss function of gold, at $k=0$. Symbols depict experimental data \cite{Palik1999} while solid lines represent the present model, together with the corresponding parameterization by Kwei \cite{Kwei1993} (dashed lines) for comparison.
	}
	\label{fig:fig1}
\end{figure}

Once the criteria for the maximum energy transfer for each inelastic interaction type are established, it is necessary to know when they apply. For this purpose, the excitation spectrum of the target (i.e., its ELF) must be examined for the different excitation types. Liquid water, an insulating material, has an ELF characterized by a single broad peak at $\sim 20$ eV \cite{Hayashi2000}, where excitations and ionizations can be separated by means of its mean binding energy $\overline{B}$ \cite{deVera2013,deVera2015}. 
The spectrum of aluminum \cite{Palik1999} is dominated by a sharp and intense plasmon excitation at $\sim 15$ eV 
\cite{Egerton2011}; thus, aluminum will be used as an example of a material where collective excitations dominate. Finally, gold and copper present complex excitation spectra where individual and collective excitations coexist. 

Figure \ref{fig:fig1} contains the optical ($k=0$) ELF of 
 gold (symbols) \cite{Palik1999}. 
The peak around 25 eV is considered to be the plasmon resonance \cite{Egerton2011}. The rest of peaks will be assigned to the excitation of the different bands. A parameterization made by Kwei \textit{et al.} \cite{Kwei1993} is shown by dashed lines. In the first part of our discussion, all excitations, except the plasmon, will be regarded as individual ones. 

We have parameterized the ELF in terms of Mermin-type (MELF) contributions \cite{Mermin1970}, introduced in Eq.(\ref{eq:IMFP}).
Threshold energies for each level have been taken from the literature \cite{Chantler2005,Kramida2018}, while fitting of the optical ELF has been constrained to respect as much as possible the number of electrons expected in each energy level, according to its electronic configuration, by evaluating individual $f$-sum rules \cite{Smith1978}. Our parameterization of the different contributions (thin solid lines in Fig. \ref{fig:fig1}) corresponds reasonably well to that by Kwei \textit{et al.} \cite{Kwei1993} (dashed lines). The thick line, representing the sum of all the contributions, agrees very well with the experimental data \cite{Palik1999}. 

\begin{figure}[t]
	\centering
\includegraphics[width=1.0\columnwidth]{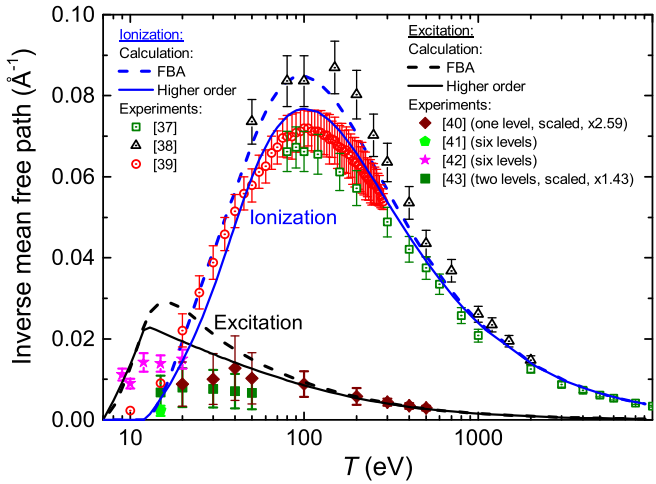}	
	\caption{(Color online) Electron inverse IMFP in water due to the processes of ionization and excitation. See the main text for the meaning of symbols and lines.
	}
	\label{fig:fig2}
\end{figure}

Our analysis of the electron IMFP in the selected materials starts with liquid water, an insulating material where both electronic transitions to bound levels (excitations) and to the conduction band (ionizations) are possible. Figure \ref{fig:fig2} depicts the inverse IMFP corresponding to these cases. 
Dashed lines represent our calculations, using $\overline{B} = 12.3$ eV, which are compared to experimental data for the water molecule scaled to liquid water density (depicted by symbols). There is plenty of experimental information for ionization \cite{Schutten1966,Bolorizadeh1986,Bull2014}, while data for excitation is rather scarce, and frequently restricted to a few excitation channels \cite{Thorn2007,Brunger2008,Ralphs2013,Matsui2016}. The latter  
have been scaled according 
to data 
 for six excitation channels \cite{Brunger2008} (the scaling factors appearing in the figure legend).
For ionization there is good agreement between our calculation and the different experiments.
For excitation, our calculation agrees with experimental data for energies $\gtrsim 40$ eV, but overestimates them below this energy, as it is expected from the FBA \cite{Emfietzoglou2013}. 

FBA results can be improved by introducing higher order corrections. By using a simple one, which accounts for the Coulomb-field felt by the primary electron in the presence of the target \cite{Salvat1985,Emfietzoglou2003}, we obtain the solid lines shown in Fig. \ref{fig:fig2}, which result in a significant improvement, especially for electronic excitations.

\begin{figure}[t]
	\centering
	\includegraphics[width=0.97\columnwidth]{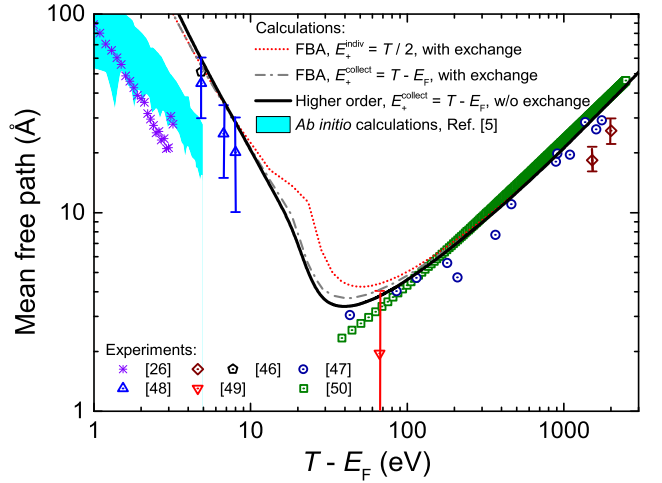}
	\caption{(Color online) Electron IMFP in Al. See the main text for the meaning of symbols, lines and the shaded area.
		}
	\label{fig:fig3}
\end{figure}

Figure \ref{fig:fig3} depicts the IMFP of electrons in aluminum, which exemplifies a material whose spectrum is dominated by one plasmon (i. e., collective) excitation. 
Symbols correspond to experimental data \cite{Bauer2015,Kanter1970,Tracy1974,Callcott1975,Powell1977,Lesiak1997}. The shaded area at low energies represents \textit{ab initio} calculations \cite{Brown2016}. 
Red dotted and gray dot-dashed lines correspond, respectively, to calculations using Eq.(\ref{eq:IMFP}) with $E_+^{\rm indiv}=T/2$ \cite{Denton2008} (individual transitions dominate) and with $E_+^{\rm collect}=T-E_{\rm F}$ \cite{Bourke2012} (collective excitations dominate); exchange is included in both cases. 
Finally, the black line uses $E_+^{\rm collect}=T-E_{\rm F}$, but excludes electronic exchange, which should not be considered for plasmon excitations due to distinguishability from the primary electron.
The latter condition
 yields the best agreement with most of the experimental data in the whole energy range, which validates the maximun energy transfer for collective excitations 
\cite{Bourke2012}, as it will be used in the following. In this case, the calculation including higher order corrections by means of the Coulomb-field perturbation term is practically identical to the black solid line.

\begin{figure}[t]
	\centering
	\includegraphics[width=0.99\columnwidth]{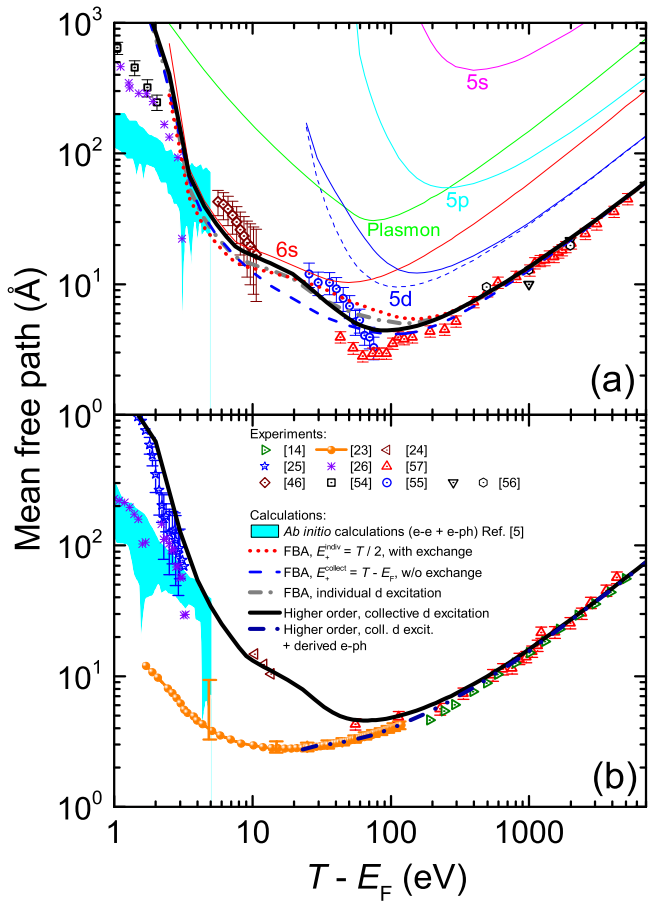}
	\caption{(Color online) Electron IMFP in (a) Au and (b) Cu.
See the main text for the meaning of symbols, lines and shaded area.
}
	\label{fig:fig4}
\end{figure}

Gold is a material with a complex excitation spectrum (Fig. \ref{fig:fig1}) where both individual and collective excitations coexist. Figure \ref{fig:fig4}(a) shows the calculated (lines) and experimental (symbols) \cite{Sze1964,Kanter1970,Lindau1976,Gergely2004,Tanuma2005,Bauer2015} electron IMFP in gold.
Thin solid lines represent contributions from different excitations, evidencing their relevance at different electron energies. The red dotted and blue dashed lines correspond, respectively, to calculations where $E_+^{\rm indiv}=T/2$ (all excitations regarded as individual, exchange included) and $E_+^{\rm collect}=T-E_{\rm F}$ (all regarded as collective, exchange not included). 
Clearly, the former overestimates the IMFP around its minimum, although its behavior at low and high energies is reasonable. Besides this, the latter also reproduces the minimum of the IMFP at $\sim 70$ eV. This manifests the importance of the collective excitations, as pointed out in Ref. \cite{Bourke2012}. However, this calculation seems to differ from the experiments at lower energies and, particularly, does not reproduce the structure of the experimental IMFP for energies $\lesssim 70$ eV, which is retrieved when all excitations are treated as individual, except for the plasmon, as 
shown by the gray dash-dotted line.

The calculated IMFP can be improved by introducing two more considerations. First, 
higher order corrections can be added as in the previous cases. Second, the excitation of the 5d electrons, with onset around 20 eV, was suggested to be an atomic giant resonance \cite{Verkhovtsev2015a,Verkhovtsev2015b}. As this is a collective atomic excitation \cite{Brechignac1994},
it has to be treated as such, and its associated IMFP evolves from the blue solid thin line to the blue dashed thin line in Fig.\ref{fig:fig4}(a). 
All the previous ingredients are incorporated in the calculations shown by the black solid thick line.
It is striking its excellent agreement with the experimental data over the entire energy range, from high energies down to 3--5 eV, 
and particularly around 5 -- 100 eV, where the structure of the experimental  IMFP is very well reproduced.

Finally, we apply the previously detailed methodology to calculate the IMFP in copper, where low energy data have recently been obtained  through XAFS experiments \cite{Bourke2010}. These measurements, shown in Fig. \ref{fig:fig4}(b) by an orange line (with symbols and error bars), are in conflict with older measurements \cite{Knapp1979,Ogawa1997,Bauer2015}.
Our full calculation (black solid line) reproduces very well the older experimental data \cite{Knapp1979,Ogawa1997,Tanuma2005,Da2014,Bauer2015} and is close to the \textit{ab initio} results from Ref. \cite{Brown2016} down to $\sim 4$ eV. 

In order to better understand the low energy discrepancy,
let us assume that other interaction mechanism might affect the 
derivation of IMFP from the XAFS data, which requires theoretical interpretation of the measurements \cite{Bourke2010}. 
The authors of Ref. \cite{Brown2016} pointed out to the role played in their \textit{ab initio} calculations by the electron-phonon interaction, quite strong in certain crystallographic orientations, resulting in a significant dispersion of their results,
which spans from the older to the XAFS-derived data.
Therefore, it is plausible that unaccounting for electron-phonon interaction in the XAFS experiment could affect the derived IMFP, $\lambda_{\rm XAFS}$.

To deeper investigate this point, we have obtained the difference $\lambda^{-1}_{\rm diff} = \lambda^{-1}_{\rm XAFS}-\lambda_{\rm e}^{-1}$, where $\lambda_{\rm e}$ is our calculated electronic IMFP. We have fitted $\lambda_{\rm diff}$ to the asymptotic form of the electron-acoustic phonon mean free path \cite{Bradford1991,Fitting2007}. The blue double-dotted-dashed curve in Fig. \ref{fig:fig4}(b) corresponds to a total IMFP calculated as $\lambda = \left( \lambda_{\rm e}^{-1}+\lambda^{-1}_{\rm diff} \right)^{-1}$, which perfectly agrees with both the high energy experimental IMFP \cite{Tanuma2005,Da2014} and the XAFS-derived IMFP \cite{Bourke2010} down to 20 eV. This indicates that electron-phonon interaction (among other possible processes) could have affected the interpretation of the XAFS experiments \cite{Bourke2010}. This plausible explanation sheds light on the disagreement between the older and the XAFS measurements at low energies.

In conclusion, we have analyzed the role played by the different excitations (collective or individual), as well as ionizations, in the maximum energy transferred 
in the inelastic interactions of an electron moving through either conducting (aluminum, gold and copper) or insulating (liquid water) media appearing in nanostructured devices and biological environments. The discussion and results presented in this work highlight the importance of a proper description of the material excitation spectrum (through its Energy-Loss Function) for an accurate calculation of the electron inelastic mean free path. Also, the need for higher order corrections to the dielectric formalism to obtain accurate inelastic mean free path at the lower energies is remarked. 
The calculated IMFP for these materials
are in excellent agreement with the experimental data in practically the entire energy range, covering from the low energies of hot electrons \cite{Bauer2015} up to the high energies typical in electron beams \cite{Egerton2011}. Our results also help to elucidate the discrepancy between recent \cite{Bourke2010} and older \cite{Knapp1979,Ogawa1997,Bauer2015} measured electron IMFP in copper at low energies, which could be due to the unaccounted electron-phonon interaction in the former. The presented results are of great relevance for understanding the dynamics of electron transport in condensed matter.

\begin{acknowledgments}
The authors are indebted to Prof. Isabel Abril for enlighting discussions and constant support while developing this work. Financial support was provided by the Spanish Ministerio de Econom\'{i}a y Competitividad and the European Regional Development Fund (Project No. FIS2014-58849-P), as well as by the Fundaci\'{o}n S\'{e}neca (Project No. 19907/GERM/15).
\end{acknowledgments}

\end{document}